\documentclass[a4paper]{mem}
\usepackage[latin1]{inputenc}
\usepackage{txfonts}
\usepackage[english]{babel}
\usepackage[dvips]{graphicx,epsfig}
\usepackage{natbib}
\begin{document}
 \title{The very young planetary nebula LS IV -12 111}
\titlerunning{The very young PN LS IV -12 111}
\author{L. Cerrigone\inst{1}, G. Umana\inst{2} and C. Trigilio\inst{2}}
\date{Granada, April 13-15 2005}
\institute{Università degli Studi di Catania, Dip. di Fisica e Astronomia - sez. Astrofisica\\
via S. Sofia, 78\\
95123 Catania, Italy\\
\email{lce@ct.astro.it}
\and 
Istituto di Radioastronomia, INAF\\
Noto VLBI Station\\
c.da Renna bassa, Noto, Italy}

\date{Granada, April 13-15, 2005}
\abstract{
In the last few years we have started a systematic study aimed at detecting very young Planetary Nebulae (YPNs). The project, based on both radio (multi-frequency and high resolution) and optical observations, has already provided interesting results, including the identification of new YPNs and the observation, for the first time in PN in this very early  phase of their formation, of radio bipolar morphologies.\\
In this paper we present radio and optical results on one of our targets, namely IRAS 19590-1249 (LS IV -12 111). The analysis of spectroscopic observations, obtained with the Anglo Australian Telescope (AAT) in different epochs, has allowed us to conclude that the optical spectrum of this object has evolved quite rapidly, showing, very recently, strong forbidden nebular emission lines consistent with a young PN. This result confirms that IRAS 19590-1249, has turned into a PN within the last decades and makes this source a perfect target for studying the early evolution and structure of Planetary Nebulae.}

\maketitle

\section{Introduction}
Planetary Nebulae (PN) are the link between the Asymptotic Giant Branch  and the White Dwarf, their envelopes resulting from the interaction between AGB ejected material and the expanding nebula. PN observations can then provide information on both the chemistry of AGB products and their diffusion into the ISM, besides the possibility to check stellar evolutionary theories.
The explanation of the often amazing structures observed in PN is in fact still a matter of discussion. In particular the timescales of the transition from the post-AGB to the PN and the origin of the shaping mechanism are not explained yet.
Important information can therefore be provided by the observation of very young PN (yPN), in which the physical processes associated with their formation are still occurring.
In the last years we have started a systematic study of a new sample of yPN, selected on the basis of their optical (B[e] spectral type) and Infrared (IRAS emission) properties.
In this paper we present the results collected so far on one of our sources, LS IV -12 111.
\section{LS IV -12 111}
Ultraviolet, optical and infrared observations of this source \citep{1992ApJ...394..298M, 1993ApJ...408..593C} indicated a B0 spectral type, making it likely to be a post-AGB star quickly evolving towards the Planetary Nebula. A comparison of its atmospherical parameters with theoretical AGB evolutionary tracks, led to estimate a 0.67 M$_\odot$ central mass \citep{1993ApJ...408..593C}. \citet{2002AstL...28..257A} have observed a photometric variability, with $\Delta m\sim0.3$ in UBV, on a 1 day timescale.
In the former papers the presence of several permitted and forbidden emission lines of Si II, [O II], [S II], [Cr II] and [Fe II] is reported. A spectral variability can be deduced by the appearance in 1994 \citep{2003A&A...401.1119R} of a 4458 Å [Fe II] line, confirmed by our 2002 spectrum.
\begin{figure}[htbp]
\centering
\includegraphics[width=6cm]{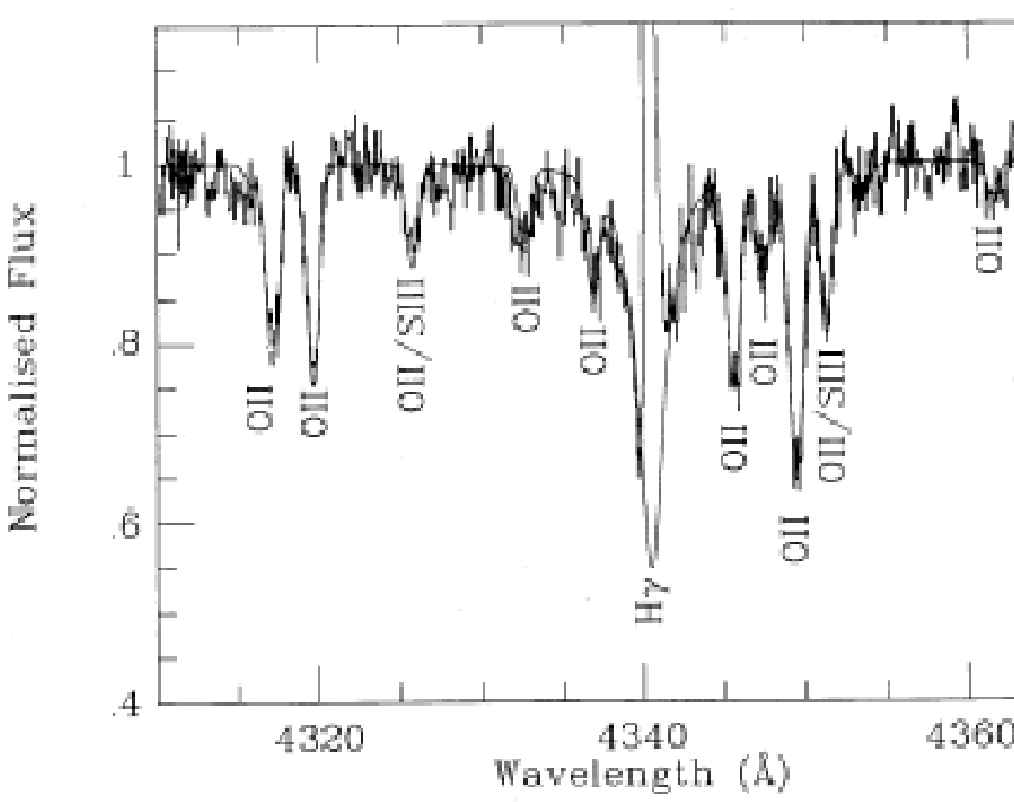}
\caption{{\footnotesize Optical spectrum of LS IV from \citet{1992ApJ...394..298M}.}}
\label{fig1}
\end{figure}\begin{figure}[htbp]
\centering
\includegraphics[width=6cm]{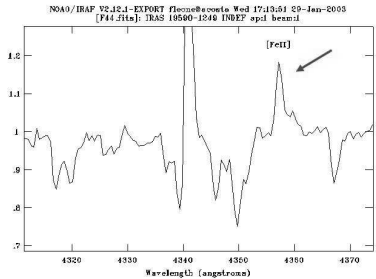}
\caption{{\footnotesize LS IV optical spectrum obtained in 2002 at AAT.}}
\label{fig2}
\end{figure}
\section{Observations and results}
We observed our source in three different epochs. The first time it was observed in June 2001 at 8.4 GHz with the Very Large Array\footnote{The VLA of the National Radio Astronomy Observatory (NRAO) is operated by Associated Universities, Inc. under a
  contract with the National Science Foundation} (VLA) in CnB configuration. The second time it was observed in February 2003 with the VLA in D array, at 4.8, 14.9 and 22.4 GHz, and then in August 2003 at 8.4 GHz in A array. Tables \ref{tab1} and \ref{tab2} show the details for the three of the runs.
\begin{table}[htbp]\scriptsize
\centering
\begin{tabular}{|lccc|}
 \hline
 \hline
 \multicolumn{4}{|c|}{February 2003} \\
 Resolution & $12''$ & $4''$ & $3''$ \\
 Flux (mJy) & $3.2\pm0.1$ & $2.3\pm0.1$ & $2.05\pm0.09$ \\
 Freq. (GHz) &$4.8$ & $14.9$ & $22.4$ \\
 \hline
 \hline
 \end{tabular}
 \caption{{\footnotesize Details about D array run; the source was point-like in this array.}}
 \label{tab1}
 \end{table}
 \vspace{-1cm}
 \begin{table}[htbp] \footnotesize \centering
 \begin{tabular}{|lcc|}\hline\hline
  & June 2001 & July 2003\\
  Resolution & $1.5''$ & $0.2''$\\
  Flux (mJy) &  $2.76\pm0.08$ & $2.69\pm0.08$\\
  Size & $\sim1.9''$ & $\sim2\times2''$\\
  \hline\hline
  \end{tabular}
  \caption{{\footnotesize Observations at 8.4 GHz were done in two different arrays (CnB in 2001 and A in 2003).}}
  \label{tab2}
  \end{table}
\section{Discussion}
\begin{figure}[htbp]
\centering
\includegraphics[width=6cm]{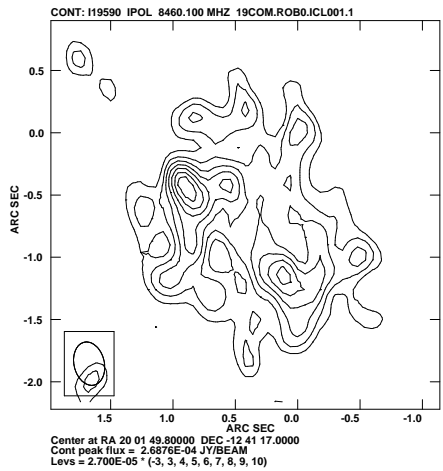}
\caption{{\footnotesize IRAS 19590-1249 high resolution map, obtained at the VLA (8.4 GHz).}}
\label{fig3}
\end{figure}

\begin{figure}[h]
\centering
\includegraphics[width=5cm]{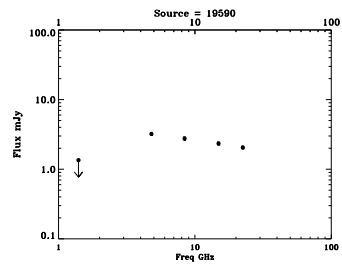}
\caption{{\footnotesize Radio spectrum of LS IV. The 1.4 GHz point is only an upper limit (3$\sigma$) from NVSS survey \citep{1998AJ....115.1693C}.}}
\label{fig4}
\end{figure}

\begin{figure}[htbp]
\centering
\includegraphics[width=5cm]{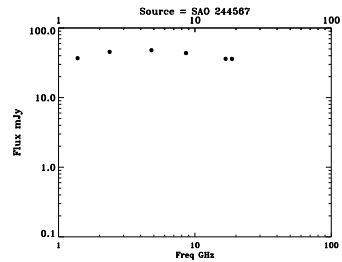}
\caption{{\footnotesize Radio spectrum of SAO 244567.}}
\label{fig5}
\end{figure}

The detection of radio flux from this star definitely defines its evolutionary status, since it is the proof of the presence of an ionized shell \citep{2004A&A...428..121U}.
Fig.\ref{fig3} lets us view for the first time the inner structure of this yPN, unveiling a bipolar morphology. This provides us with a strong constrain on PN evolutionary models, since it implies that, whatever its origin, the shaping mechanism must have been active well before the onset of ionization, which, as witnessed by spectral variations, is still an ongoing process in our source.
Moreover, current PN evolutionary models consider that, because of their compactness and then high density, young PN should be optically thick in radio wavelengths even at high frequencies (i.e., above 5 GHz) \citep{2000kwok}. The critical frequency, at which the nebula becomes optically thin, could then be used as an age indicator \citep{1991ApJ...378..599A}.
Yet, the radio spectrum we built up for LS IV -12 111 (Fig.\ref{fig4}) does not seem to confirm this pattern, although optical spectroscopic variations and Far Infrared excess (IRAS) let us think this PN formed in the last decades.
\citet{2004ASPC..313..134T} have performed, with the Australia Telescope Compact Array\footnote{Australia Telescope Compact Array is funded by the Commonwealth of Australia for operation as a National Facility managed by CSIRO}, multi-frequency observations of the youngest PN known so far, SAO 244567. We can therefore compare this source's spectrum (Fig.\ref{fig5}) to LS IV's one.
This comparison shows us that SAO 244567 and LS IV -12 111, though being both very young, are optically thin at 5 GHz, with the latter source likely to have a sharp flux decrease between 1 and 4 GHz. Such a behavior of radio spectra is not in agreement with current models and probably needs further investigation.

\end{document}